\documentclass[aps,reprint,twocolumn,showpacs,amsmath,amssymb, 
superscriptaddress]
{revtex4-2}
\usepackage{graphicx}
\usepackage{color}
\usepackage[ 
colorlinks=true,
pdfborder={0 0 0},
breaklinks,
allcolors=blue
]{hyperref}
\usepackage{orcidlink}

\begin{document}
\title{Secondary-Structure Phase Formation for Semiflexible 
Polymers\\ by Bifurcation in Hyperphase Space}
\author{Dilimulati Aierken\,\orcidlink{0000-0003-1727-5759}}
\email{d.aierken@princeton.edu}
\affiliation{Department of Chemical and Biological Engineering, Princeton 
University, Princeton, NJ 08544, USA}
\affiliation{Omenn-Darling Bioengineering Institute, Princeton University,
Princeton, NJ 08540, USA}
\affiliation{Soft Matter Systems Research Group, Center for Simulational
Physics, Department of Physics and Astronomy, The University of Georgia, 
Athens, GA 30602, USA}
\author{Michael Bachmann}
\email{bachmann@smsyslab.org; https://www.smsyslab.org}
\affiliation{Soft Matter Systems Research Group, Center for Simulational
Physics, Department of Physics and Astronomy, The University of Georgia, 
Athens, GA 30602, USA}
\begin{abstract}
Canonical analysis has long been the primary analysis method for studies of
phase transitions. However, this approach is not sensitive
enough if transition signals are too close in temperature space.  
The recently introduced generalized microcanonical inflection-point analysis 
method not only enables the systematic identification and classification of 
transitions in systems of any size, but it can also
distinguish transitions that standard canonical analysis cannot resolve.
By applying this method to a generic coarse-grained model for  
semiflexible polymers, we identify a mixed structural phase dominated by 
secondary structures such as hairpins and loops that originates from a 
bifurcation in the hyperspace spanned by inverse temperature and bending 
stiffness. This intermediate phase, which is embraced by the 
well-known random-coil and toroidal phases, is testimony to the necessity of 
balancing entropic variability and energetic stability in functional 
macromolecules under physiological conditions.
\end{abstract}
\maketitle 
\section{Introduction}
In recent years, substantial research interest has been dedicated to
applications in microbiology and nanotechnology on microscopic and mesoscopic
scales, where surface effects can not be ignored. Different aspects of
biomolecules have been studied extensively. As biomolecules fold into specific
structures to perform biological functions in living cells, the computational
modeling of these biopolymers, with the advantage of more precise control
compared to experiments, has been a crucial way to study structural  
transitions, which furthermore leads to applications in many areas, e.g., 
drug discovery and design~\cite{ferreira2015,gao2021,schmidt2014,smiatek2020}.
Depending on the objective, biopolymer models with different 
degrees of complexity have been introduced. All-atom simulations provide
high resolution for studies of local structure dynamics, but this comes at a 
very high computational cost and only a limited 
timescale can be covered~\cite{kmiecik2016,singh2019}.
In addition, all-atom simulations often require $\mathcal{O} (10^3)$ 
mostly empirical ``force
field'' parameters. 

On the other hand, coarse-grained models enable extensive studies of polymer 
systems at far less computational cost as less relevant 
degrees of freedom are integrated out. The underlying 
atomic interactions are replaced by effective interactions between 
monomers. Extending the coarse-graining procedure further, a monomer can also 
represent an entire chemical group of atoms or even sections of repeating 
structures or subunits. Moreover, coarse-grained modeling allows for the 
systematic study of specific aspects and thus provides a generic insight into 
the macroscopic properties that are not limited to specific 
biomolecules~\cite{bachmann2014}. This is the approach we pursue in this 
study.

DNA, RNA, and proteins can be considered semiflexible polymers, 
which are characterized by their bending stiffness or finite persistence 
length. Moreover, bending stiffness has a sigificant impact on biological 
functions and processes. It helps DNA pack in an organized way for efficient 
translation and transcription processes~\cite{yuan2008}. This is important 
as the length of DNA is very long compared to the size of the cell 
nucleus it resides in. It is also known that RNA stiffness can affect the 
self-assembly of virus particles~\cite{li2018}.  

One of the simplest semiflexible polymer models is the well-known 
Kratky-Porod or wormlike-chain model~\cite{kratky1949}, which has been 
successfully used in studies of structural and dynamic properties of 
semiflexible polymers. However, the lack of self-interactions in this model 
does not allow for the study of structural phase transitions. Therefore, 
coarse-grained polymers models with monomer-monomer interaction have been 
employed to study the phase behavior of semiflexible
polymers~\cite{seaton2013,marenz2016,skrbic2016,wu2018,wu2018b,ab1,%
majumder2021,walker2022,shakirov2023,aierken2023,ab3}, usually by means of 
conventional canonical 
statistical analysis. In this context, it is important to note that 
biological systems are finite in nature and finite-size scaling is not a
generally applicable approach to study the structural phase behavior of these 
systems. Also,
results obtained by canonical statistical analysis are often inconsistent and
ambiguous for finite systems. Therefore, it is certainly beneficial to explore 
other approaches. One candidate is the recently introduced generalized 
microcanonical inflection-point analysis method that 
provides a 
systematic and consistent approach to phase transitions in systems of any 
size~\cite{qi2018}.

In this paper, we employ this statistical analysis 
method to systematically investigate structural transitions for a generic 
coarse-grained semiflexible polymer model~\cite{aierken2023}.
Section~\ref{sec:mod_meth} describes the 
model, the simulation techniques, and the microcanonical inflection-point 
analysis
method. This is followed by the discussion and comparison of canonical and 
microcanonical results in Section~\ref{sec:analysis}. Finally, the paper is 
concluded by a summary in Section~\ref{sec:sum}.
\section{Models and Methods}
\label{sec:mod_meth}
\subsection{Coarse-grained Semiflexible Polymer Model}
For our study, we use a generic, self-interacting semiflexible homopolymer 
model. The total
energy of a polymer with $N$ monomers in a conformation 
$\boldsymbol{X}=\left(\boldsymbol{r}_1,...,\boldsymbol{r}_N\right)$, where
$\boldsymbol{r}_n$ is the position vector of the $n$th monomer, is composed of
contributions from bonded and nonbonded interactions between 
monomers, along
with an energetic bending penalty:
\begin{equation}
E(\boldsymbol{X})=\sum_{n}V_{\mathrm{b}}(r_{n\,n+1})+
\sum_{n<m+1}V_{\mathrm{nb}}(r_{nm})+
\sum_{k}V_{\mathrm{bend}}(\theta_{k}).
\label{eq:hamiltonian}
\end{equation}
Here $r_{nm}=|\boldsymbol{r}_n-\boldsymbol{r}_m|$ is the distance between
monomers $n$ and $m$, and $\theta_{k}$ is the bond angle between two adjacent
bonds.

For the interactions between nonbonded monomers, we employ the standard 12-6
Lennard-Jones (LJ) potential
\begin{equation}
V_{\mathrm{LJ}}(r)=4\epsilon_{\mathrm{LJ}}
\left[\left(\dfrac{\sigma}{r}\right)^{12}-
\left(\dfrac{\sigma}{r}\right)^{6}\right],
\label{eq:lj_potential}
\end{equation}
where $r$ is the monomer-monomer distance, $\sigma=2^{-1/6}r_{0}$ is the van
der Waals distance associated with the potential minimum at $r_0$, and
$\epsilon_{\mathrm{LJ}}$ is the energy scale. For computational efficiency, we
introduce a cutoff at $r_c=2.5\sigma$. Shifting the potential by the 
constant
$V_{\mathrm{shift}}\equiv V_{\mathrm{LJ}}(r_c)$ avoids a discontinuity  
at $r_c$.
Thus, the potential energy of nonbonded monomers is given by
\begin{equation}
V_{\mathrm{nb}}(r)=
\begin{cases}
V_{\mathrm{LJ}}(r)-V_{\mathrm{shift}}, & r<r_c,          \\
0,                                     & \text{otherwise.}
\end{cases}
\end{equation}
The interaction between bonded monomers is 
modeled by a combination of the Lennard-Jones potential and the finitely 
extensible nonlinear
elastic (FENE) potential:
\begin{equation}
V_{\mathrm{b}}(r)=V_{\mathrm{FENE}}(r)+V_{\mathrm{LJ}}(r)-V_{\mathrm{shift}},
\end{equation}
where the same parameter values of $V_{\mathrm{LJ}}(r)$ are chosen as for
nonbonded monomer-monomer interactions. The FENE potential is given
by~\cite{bird1987,kremer1990,milchev2001}:
\begin{equation}
V_{\mathrm{FENE}}(r) = -\frac{1}{2}KR^{2}\mathrm{ln}
\left[1-\left(\frac{r-r_{0}}{R}\right)^{2}\right].
\label{eq:FENE}
\end{equation}
The parameters are fixed to standard values $R=(3/7)r_{0}$ and
$K=(98/5)\epsilon_{\mathrm{LJ}}/r_0^2$~\cite{qi2014} and the
bond length $r$ is restricted to fluctuations within the range $[r_{0}-R,
r_{0}+R]$. With these parameters, the minimum of $V_{\mathrm{b}}$ is located 
at $r_0$.

To model bending strength, an additional standard potential
is introduced. Any deviation of bond angle $\theta$ from the reference angle
$\theta _{0}$ between neighboring bonds is subject to an energy penalty of the
form:
\begin{equation}
V_{\mathrm{bend}}(\theta)=\kappa\left[1-\cos(\theta-\theta_{0})\right].
\end{equation}
The parameter $\kappa \geq 0$ controls the stiffness of the polymer chain. For
$\kappa=0$, the model describes flexible
polymers~\cite{qi2019}. In this study, we chose
$\theta_{0}=0$. Therefore, any deviation from the straight chain is 
energetically unfavorable for $\kappa > 0$.

In simulations and statistical analysis of the results, we set 
$k_\mathrm{B}=1$
(Boltzmann constant), $\epsilon_{\mathrm{LJ}} = 1$, and $r_{0}=1$. The 
flexible
chain with $N=55$ monomers has already been studied extensively in the
past~\cite{qi2019}
and serves as the reference for the comparison with the semiflexible  
model. This
chain length is sufficiently short to recognize finite-size effects but long
enough for the polymer to form stable phases. The results we obtained for 
this polymer have also been verified for chains with up to $100$ 
monomers.
\subsection{Replica-exchange Simulations of Semiflexible Polymers}
The density of states of a system is a core quantity for the microcanonical
inflection-point analysis. However, its
precise estimation is challenging for complex systems, even with modern 
computer systems and advanced simulation techniques. For our study, we  
employed an extended version of replica-exchange Monte
Carlo (parallel 
tempering)~\cite{swendsen1986,ferrenberg1988,geyer1991,hukushima1996,%
hukushima1996b,earl2005}. In contrast to simulating the system at fixed 
temperature, as it is done in conventional Metropolis sampling, parallel 
tempering has been shown
to reach equilibrium faster in
simulations at low 
temperatures~\cite{fiore2008,kofke2004,machta2009,machta2011,predescu2005}.
Parallel tempering is a generalized-ensemble method; the 
microstate probability distribution is governed by the product of 
Boltzmann factors at all simulation temperatures. It samples the entire state 
space much more effectively than an ordinary Metropolis 
Monte Carlo method, which simulates an actual canonical ensemble at a single 
simulation temperature. The improved performance of parallel tempering is 
achieved by occasional exchanges of the conformations (replicas) between 
Metropolis simulation threads running at different temperatures.

Replicas of the system are simulated at different 
inverse
thermal energies $\beta_i \in [\beta_{{\min}}, 
\beta_{{\max}}]$, with $i =
1,2,\dots, I$,  where $I$ is the total number of threads. Here, 
$\beta=1/k_\mathrm{B}T_\mathrm{can}$, where $T_\mathrm{can}$ is the 
canonical heat-bath temperature. One
obvious advantage of this algorithm is that it can simulate the system under 
different conditions simultaneously. 
At each temperature, Metropolis sampling 
is performed with the acceptance probability:
\begin{equation}
P(\boldsymbol{X}\to
\boldsymbol{X}') = {\min} \left[1, 
\sigma(\boldsymbol{X},\boldsymbol{X}')
\omega(\boldsymbol{X},\boldsymbol{X}')\right],
\label{eq:metropolis_general}
\end{equation}
where $\omega(\boldsymbol{X},\boldsymbol{X}') = 
\exp\{-\beta[E(\boldsymbol{X}')-E(\boldsymbol{X})]\}$ and 
$\sigma(\boldsymbol{X},\boldsymbol{X}')$ is the ratio of forward and backward 
selection probabilities, which are usually not identical for composite 
moves such as bond-exchange updates~\cite{schnabel2011b}.
We also used displacement updates~\cite{williams2016}, crankshaft
rotations~\cite{austin2018a}, and pivot rotations~\cite{bachmann2014} to 
alter polymer conformations.
\begin{figure}
\centering
\includegraphics[width=\columnwidth]{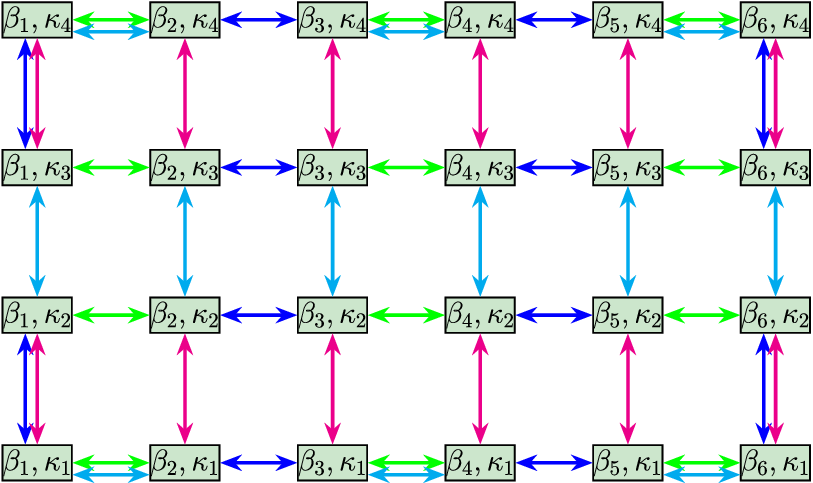}
\caption{Illustration of the extended replica-exchange Monte Carlo 
scheme in the combined parameter space of six inverse temperatures 
$\beta_1,\ldots,\beta_6$ and four bending stiffness values 
$\kappa_1,\ldots,\kappa_4$. Each node $(i,j)$ represents a simulation thread 
with a parameter combination of $(\beta_i, \kappa_j)$. The four 
exchange directions, which are chosen randomly from a uniform distribution, 
are colored differently. The additional exchanges at edge and corner nodes 
provide for the conservation of replica flows. In our actual simulations, 
typical grids contained $60\times 4$ parameters in this space.}
\label{fig:2d_sketch}
\end{figure}

In our parallel tempering simulations an 
exchange between the replica with conformation
$\boldsymbol{X}$ at inverse temperature $\beta_i$ and the replica in state
$\boldsymbol{X}'$ at $\beta_j$ is proposed after $3000$ 
sweeps (here a sweep corresponds to $N=55$ Monte Carlo updates). The exchange 
probability is
given by
\begin{equation}
a(\boldsymbol{X}\leftrightarrow
\boldsymbol{X}';\beta_i, \beta_j) = {\min} 
\left(1, e^{(\beta_i - \beta_j)(E(\boldsymbol{X})-E(\boldsymbol{X}'))}  
\right).
\label{eq:PT}
\end{equation}
For the selection of temperature sets that enable efficient exchange between 
replicas, we found the combination of the
geometric and energy methods~\cite{hukushima1999,rozada2019} most reasonable. 
In these methods, short runs with
geometric temperature sets at fixed temperature limits $\beta_{\min}$,
$\beta_{\max}$, and number of threads $I$ are performed. The
temperature for the simulation thread $i$ is initialized as $\beta_i =
\beta_{\min} G^{i-1}$, where 
$G =\left(\beta_{\max}/\beta_{\min}\right)^{1/(I-1)}$. 
Then the temperatures are adjusted by inserting the estimated average energies 
in Eq.~(\ref{eq:PT}) to get uniform expected exchange probabilities (among 
all temperature points).
For this study, we
used the following parameters: $\beta_{\min} = 0.25$, $\beta_{\max} = 50$,  
and $I=60$. In typical simulations, up to $10^9$ sweeps were performed.
\begin{figure}
\centering
\includegraphics[width=1.0\linewidth]{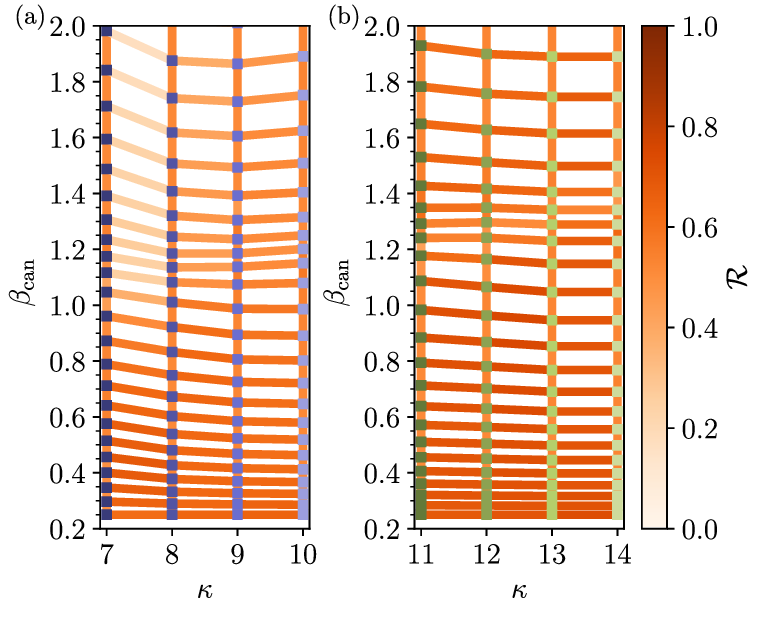}
\caption{Color-coded replica-exchange acceptance rates of
parallel tempering simulations in the combined space of inverse
temperature and bending stiffness 
for (a) for $\kappa = 7,8,9,10$ and (b) for $\kappa = 11,12,13,14$ between 
simulation threads (squares) in parameter space. The darker the line color 
the higher the acceptance rate.}
\label{fig:2d-k10-14}
\end{figure}

Even though advanced updates were used in our implementation of parallel 
tempering, the simulation dynamics at low temperatures was still too 
slow and equilibrium rarely reached. Therefore, we expanded 
the exchange of replicas 
to
the combined parameter space of simulation temperature and bending stiffness
for increased 
efficiency~\cite{marenz2016,majumder2021}.
For this purpose, the system energy is decoupled,
\begin{equation}
E(\boldsymbol{X}) = E_0(\boldsymbol{X})+\kappa E_1(\boldsymbol{X}),
\end{equation}
where $E_0(\boldsymbol{X}) = \sum_{n}V_{\mathrm{b}}(r_{n\,n+1}) +
\sum_{n<m+1}V_{\mathrm{nb}}(r_{nm})$ and
$E_1(\boldsymbol{X})=\sum_{k}\left[1-\cos\theta_k\right]$. 
Consequently, the exchange probability of replica $\boldsymbol{X}$ with 
bending
stiffness $\kappa_i$ at inverse temperature $\beta_i$ and replica
$\boldsymbol{X}'$ with bending stiffness $\kappa_j$ at inverse temperature
$\beta_j$ is given by 
\begin{equation}
P_{\mathrm{ext}}=\mathrm{min}\left(\exp \left[ (\Delta \beta \Delta E_0)
+\Delta(\beta \kappa)\Delta E_1 \right], 1 \right).
\end{equation}
Here $\Delta \beta = \beta_i-\beta_{i+1}$ and $\Delta(\beta \kappa)=
\beta_{i}\kappa_{i}-\beta_{i+1}\kappa_{i+1}$. The simulation setup is
illustrated in Fig.~\ref{fig:2d_sketch}.

In this study, we found $\Delta \kappa = 1$ is a sufficient spacing for the
range of $\kappa$ values studied. Additional intermediate $\kappa$ values were
added, where a finer resolution was needed. In this region of bending 
stiffnesses,
we first simulated two sets of four $\kappa$ values each, $\{7, 8, 9, 10\}$ 
and
$\{11, 12, 13, 14\}$, where $60$ $\kappa-$dependent temperatures, 
adjusted for
parallel tempering, are used for each $\kappa$ value. On occasion, we 
then also included additional $\kappa$ values to resolve more details of 
phase behavior in particularly interesting regions of the $\beta-\kappa$ 
hyperphase diagram.

Combining the energy histograms obtained at different temperatures at fixed 
$\kappa$ values, the multi-histogram
reweighting
method~\cite{ferrenberg1988,kumar1992a}
yielded an improved estimator for the density of states $g(E)$ that covers the
entire energy range at fixed bending stiffness. For further analysis, the 
B\'ezier
method~\cite{bezier1968,gordon1974,bachmann2014}
was used to smooth the microcanonical entropy curves $S(E)=k_\mathrm{B}\ln 
g(E)$ and to calculate its derivatives in preparation of the subsequent
microcanonical inflection-point analysis.

In order to assess the performance of the extended parallel-tempering simulation method, we
estimated the replica-exchange acceptance rates $\mathcal{R}$. The results are shown in
Fig.~\ref{fig:2d-k10-14} for two different $\kappa$ ranges. The red color shade encodes the value of
$\mathcal{R}$. As we see, all pairs of connected simulation threads (marked by squares) exhibit very
good replica-exchange behavior. For the most part, the extensive parameter tuning for
$\beta_\mathrm{can}$ kept exchange rates in the optimal $0.3<\mathcal{R}<0.7$ range. However, we
also observe distinct differences. Whereas the exchange rates are rather uniform for the larger
$\kappa$ set as shown in Fig~\ref{fig:2d-k10-14}(b), the obviously reduced exchange rates between
$\kappa = 7$ and $\kappa = 8$ for inverse temperatures $\beta_{\mathrm{can}} > 1.1$ suggest that the
system behavior might qualitatively change for $\kappa = 7$ and $\kappa = 8$, hinting at a possible
phase transition in this region of parameter space. This will be investigated in greater detail
later by means of microcanonical inflection-point analysis.

\begin{figure}
\centering
\includegraphics[width=1.0\linewidth]{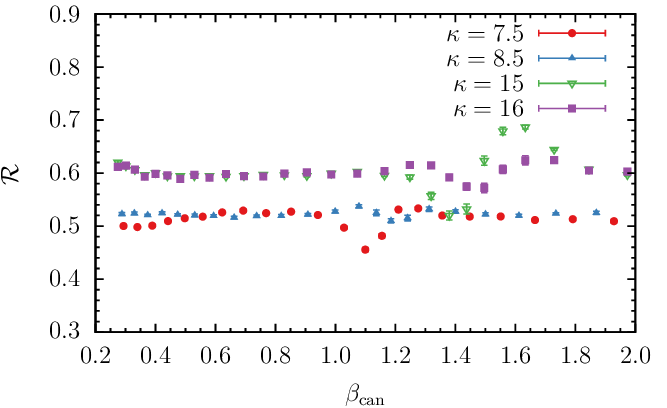}
\caption{Measured replica-exchange acceptance rates
for $\kappa = 7.5, 8.5, 15, 16$ in inverse-temperature space.
\label{fig:newRate}}
\end{figure}

Figure~\ref{fig:newRate} shows the quantitative inverse-temperature 
dependence of $\mathcal{R}$ for selected $\kappa$ values. Again, carefully 
choosing the inverse simulation temperatures $\beta_\mathrm{can}$ establishes 
almost uniform exchange rates as desired. However, we also observe a major 
fluctuation in all cases, which shifts toward higher inverse temperatures as 
$\kappa$ is increased, from $\beta_{\mathrm{can}} \approx 1.15$ for $\kappa =
7.5$ to $\beta_{\mathrm{can}} \approx 1.6$ for $\kappa = 16$. Typically, 
large deviations in the exchange rates from a uniform distribution can be 
associated with phase transitions, which, as we will see, is indeed the case 
here, too. 
\subsection{Generalized Microcanonical Inflection-point Analysis}
The generalized inflection-point analysis
method~\cite{qi2018} for the systematic
identification and classification of transitions in systems of any size
combines microcanonical
thermodynamics~\cite{gross2001} and the
principle of minimal
sensitivity~\cite{stevenson1981,stevenson1981b}. It
has already led to novel insights into the nature of phase transitions. Even
the Ising model, which has been excessively studied for almost a century,
possesses a more complex phase structure than previously
known~\cite{sitarachu2020,sitarachu2020b,sitarachu2022}. This method has also 
been employed
to particle aggregation~\cite{trugilho2022},
self-assembly kinetics in macromolecular
systems~\cite{trugilho2022b}, and in studies of the
general geometric and topological foundation of transitions in phase
space~\cite{bel-hadj-aissa2020,gori2022,pettini2019,dicairano2022}.
It motivated the further investigation of higher-order derivatives of the
Boltzmann microcanonical entropy with an additional conserved
quantity~\cite{bel-hadj-aissa2020b} and has even been used as
justification for pattern recognition criteria in computer
science~\cite{chaudhuri2021}.

It is important to note that the attribute ``microcanonical'' has only 
historical reasons. The analysis method used here has nothing to do with the 
microcanonical ensemble. The system energy $E$ is not constant. In fact, the 
density of states, $g(E)$, is a function of energy and it can also be used to 
calculate energetic ``canonical'' averages.

Microcanonical statistical analysis is based on the assumption that 
entropy $S$ and system energy $E$
control the phase behavior of any system. The microcanonical Boltzmann
entropy 
$S(E)=k_{\mathrm{B}}\ln g(E)$ relates these quantities to each other. If the 
system does not experience phase transitions, the entropy curve
$S(E)$ and its derivatives exhibit well-defined concave or convex monotony.
However, phase transitions in the system will alter the monotonic behavior, 
even for finite systems. 

From the canonical statistical analysis of first- and
second-order transitions, it is known that entropy and/or internal energy 
$\langle E\rangle$
rapidly change (or are most sensitive), if the heat-bath temperature 
$T_\mathrm{can}$ is varied near the
transition point. It can also be interpreted as the temperature change being
\emph{least} sensitive to a change of the internal energy near the transition 
point.

This behavior corresponds to the least-sensitive dependency of microcanonical
quantities in the space of system energies. Thus, a phase transition causes a
least-sensitive inflection point in the microcanonical entropy or its 
derivatives. Therefore, an inflection
point can be associated with an extremum in the next-higher derivative at the
transition energy, which simplifies the precise identification of the 
transition point. By systematically analyzing these alterations, different 
types of
transitions can be classified.
\begin{figure}
\centering
\includegraphics[width = 1.0\columnwidth]{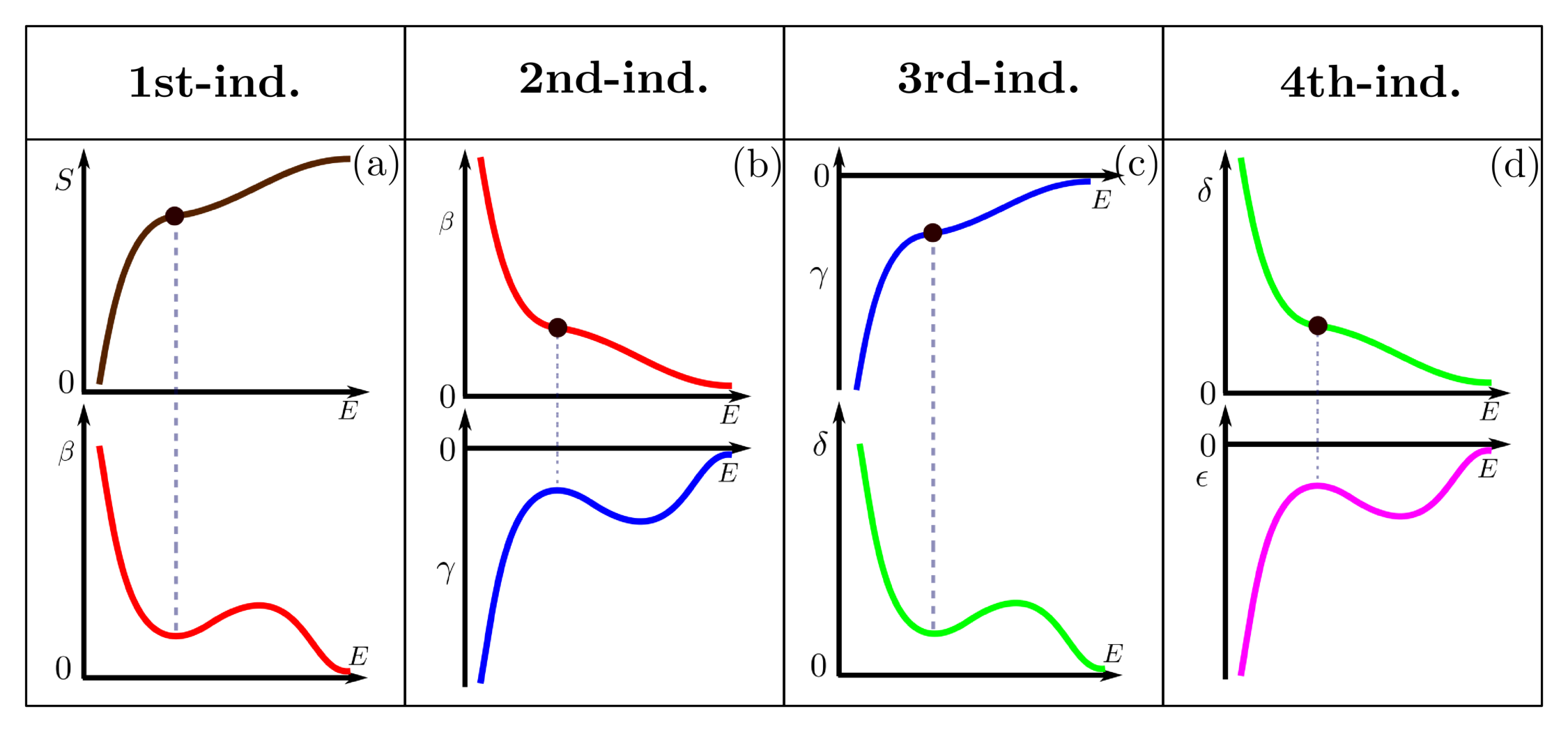}
\caption{Sketch of \emph{independent transitions} up to fourth order as
defined in the microcanonical inflection-point analysis method. (a) A
first-order independent transition is characterized by a least-sensitive 
inflection point in
$S(E)$, which corresponds to a positive minimum in $\beta = dS/dE$. (b) For a
second-order transition, the inflection point in $\beta$ is
associated with a negative maximum in $\gamma = d^2S/dE^2$. (c) An
inflection point in $\gamma$ defines a third-order transition
and $\delta=d^3S/dE^3$ exhibits a positive minimum. (d) A fourth-order
transition possesses an inflection point in $\delta$ and
$\epsilon = d^4 S/dE^4$ a negative maximum.}
\label{fig:micro_independent}
\end{figure}
\begin{figure*}
\centering
\includegraphics[width=0.8\linewidth]{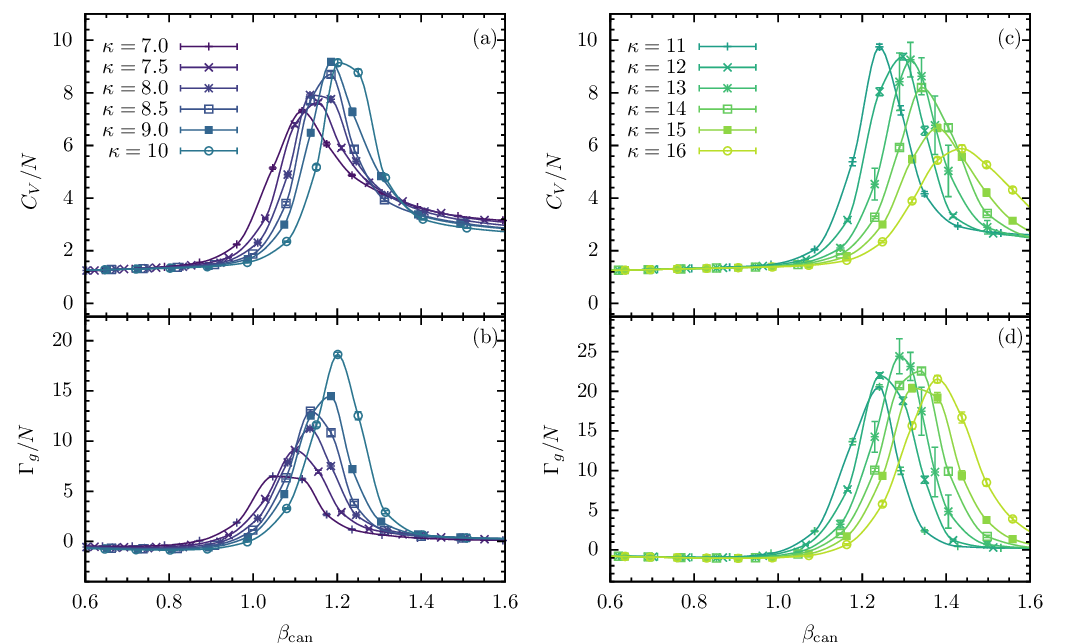}
\caption{Thermal fluctuations of (a) energy (heat capacity
$C_V=d\langle E\rangle/dT_\mathrm{can}$) and (b)
square radius of gyration 
($\Gamma_g = d\langle R^2_\mathrm{gyr}\rangle/dT_\mathrm{can}$), 
plotted as functions of $\beta_\mathrm{can}$ at selected values of the 
bending stiffness in the range $7\leq \kappa \leq 10$. (c), (d) Same for 
examples in the interval $11\leq \kappa \leq 16$.}
\label{fig:2d-k7-16:can}
\end{figure*}

In this scheme, a first-order transition in $S(E)$ is signaled by a
least-sensitive inflection point at transition energy $E_{\mathrm{tr}}$. 
Therefore, the
first derivative, i.e., the inverse microcanonical temperature $\beta(E)
\equiv dS/dE$, forms a backbending region as shown in
Fig.~\ref{fig:micro_independent}(a) that possesses a positive-valued 
minimum in $\beta(E)$ at $E_\mathrm{tr}$,
\begin{equation}
\beta(E_\mathrm{tr})=\dfrac{d
S(E)}{dE}\bigg|_{E=E_\mathrm{tr}}>0.
\end{equation}
Similarly, if there is a least-sensitive inflection point in
$\beta(E)$, the phase transition is classified as a second-order transition.
As
shown in Fig.~\ref{fig:micro_independent}(b), the derivative of $\beta(E)$ has
a negative-valued peak at the transition energy $E_\mathrm{tr}$,
\begin{equation}
\gamma(E_\mathrm{tr})=\dfrac{d^2 S(E)}{d
E^2}\bigg|_{E=E_\mathrm{tr}}<0.
\end{equation}
This approach can be generalized to any order. Therefore, for a
transition of odd order $(2k-1)$ ($k$ is a positive integer), the $(2k-1)$th
derivative of $S(E)$ possesses a positive-valued minimum,
\begin{equation}
\dfrac{d^{(2k-1)} S(E)}{dE^{(2k-1)}}\bigg|_{E=E_\mathrm{tr}}>0,
\end{equation}
and a transition of even order $2k$ is characterized by a
negative-valued peak in the $2k$th derivative,
\begin{equation}
\dfrac{d^{2k} S(E)}{d E^{2k}}\bigg|_{E=E_\mathrm{tr}}<0.
\end{equation}
We call this transition type an 
$\textit{independent}$ phase 
transition. These turn into the known thermodynamic phase 
transitions in the thermodynamic limit of infinitely large systems. 
However, this terminology implies that there is another transition type,
$\textit{dependent}$ transitions~\cite{qi2018}. A dependent transition is 
inevitably associated with an independent transition and it is located in the 
disordered phase closest to the 
independent transition. Therefore a dependent transition can be interpreted 
as precursor of the independent transition it coexists with. Although 
dependent transitions are less common
than independent transitions (only few independent transitions seem to have 
a dependent companion), they can provide valuable insights into the general
nature of transition behavior in complex systems. In the microcanonical 
inflection-point study of the two-dimensional
Ising model, a third-order dependent transition, associated with the 
well-known critical transition, was
identified~\cite{sitarachu2022}. It was found to be caused by a
collective preordering of spins in the paramagnetic phase. In our
study of the phase behavior of semiflexible polymers, dependent transitions  
were not found, though.
\section{Bifurcation in the Hyperphase Diagram of Semiflexible Polymers}
\label{sec:analysis}
In the following we analyze the transition behavior of semiflexible polymers 
from both canonical and microcanonical perspectives. For this purpose we 
distinguish the canonical inverse heat-bath temperature 
$\beta_{\mathrm{can}}$ as a thermodynamic state variable from the 
microcanonical temperature $\beta(E)$, which is a system property.
\subsubsection{Canonical Statistical Analysis}
Canonical response quantities, such as heat capacity $C_V=d\langle
E\rangle/dT_\mathrm{can}$ and fluctuations of the square radius of gyration,
$\Gamma_g = d\langle R^2_\mathrm{gyr}\rangle/dT_\mathrm{can}$, are shown in 
Fig.~\ref{fig:2d-k7-16:can} as
functions of $\beta_\mathrm{can}$ for a broad 
range
of bending parameter values. For $7 \leq \kappa \leq 10$ 
[Fig.~\ref{fig:2d-k7-16:can}(a)], we find that 
peak values of $C_V$ increase for larger bending stiffness. The 
peak locations shift to larger inverse transition temperatures. The same 
trend is observed for the peaks in the fluctuations of the square radius of
gyration, shown in Fig.~\ref{fig:2d-k7-16:can}(b). In fact, these signals 
are extrapolations of the $\Theta$ collapse transition, well-known from 
flexible polymers ($\kappa=0$), into the nonzero $\kappa$ regime. 

However, surprisingly, a different trend is observed for the heat capacities 
in the range $11 \leq \kappa \leq 16$ [Fig.~\ref{fig:2d-k7-16:can}(c)]. The 
peak values consistently decrease and the peaks broaden again for larger
$\kappa$ values. The peak values of the fluctuations of the radius gyration 
do not keep increasing either. It is important to note that only one major
peak is found in these quantities that still suggests a single collapse 
transition with
enhanced thermal activity between entropically favored wormlike chains at
higher temperatures and energetically more ordered structures at lower
temperatures. As it will turn out later in the structural analysis, this is 
an incomplete interpretation. Canonical statistical analysis averages out 
vital information.

Also note the typical ambiguity in the canonical analysis for
finite systems. As it can be seen in Fig.~\ref{fig:2d-k7-16:can}, 
peaks in the energetic and
structural fluctuations locate the transition points at different
temperatures at any given $\kappa$ value. Therefore, in the following, we 
employ a different analysis approach that helps dissolve the 
canonical ambiguities in this interesting parameter range for semiflexible 
polymers and provides a quite clear picture of the actual transition behavior 
of the system. 
\subsubsection{Microcanonical Inflection-Point Analysis}
\begin{figure}
\centering
\includegraphics[width=0.5\textwidth]{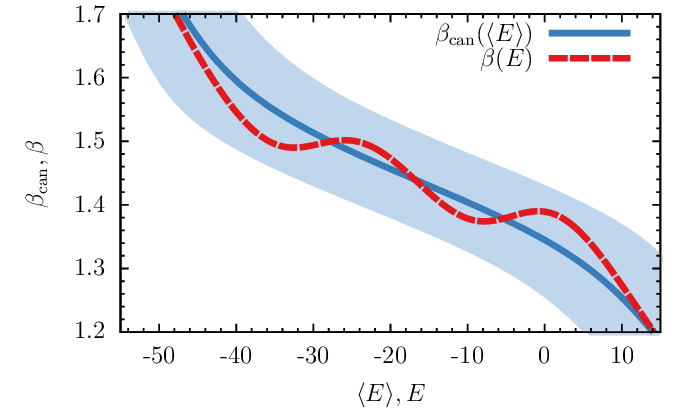}
\caption{Canonical and microcanonical results for the relationship between 
inverse temperatures  
and energies [$\beta_\mathrm{can}(\langle E\rangle)$ and $\beta(E)$, 
respectively] for
the semiflexible 55-mer at $\kappa=16$. The shaded area is the 
standard deviation of the system energy $\sigma(E)$, which represents the 
thermal fluctuations of the system energy $E$ at the corresponding inverse 
heat-bath temperature $\beta_\mathrm{can}$.}
\label{fig:comp}
\end{figure}

Figure~\ref{fig:comp} nicely illustrates the relationship between inverse 
temperature and energy in both canonical and microcanonical statistical 
analysis. The canonical expectation value of the energy 
$\langle E\rangle(\beta_\mathrm{can})$ 
averages out all fluctuations in system energy 
$E$ and $\beta_\mathrm{can}(\langle E\rangle)$ (blue solid curve)
exhibits only a single inflection point, indicating a single transition. In 
contrast, the microcanonical curve $\beta(E)$ (red dashed curve) reveals 
two (first-order) transition signals instead in this energy region. The 
shaded area represents the canonical standard deviation or fluctuation range 
(which corresponds to the heat capacity). It completely consumes 
the hierarchical microcanonical transition signals and, therefore, we 
conclude that the canonical statistical approach is not sufficiently sensitive 
for the analysis of intricate transition behavior in finite systems.

\begin{figure}
\centering
\includegraphics[width=0.5\textwidth]{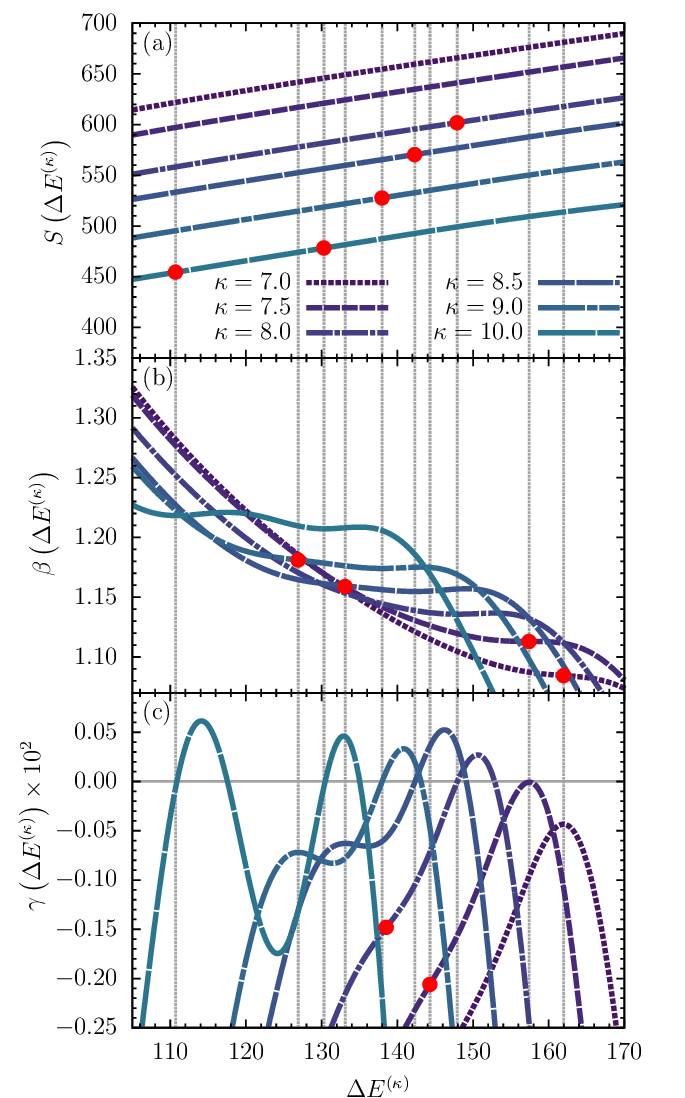}
\caption{(a) Microcanonical entropy $S$ and its derivatives (b) $\beta =
dS/dE$, and (c) $\gamma = d\beta/dE$ for semiflexible polymers with
$\kappa=7.0,\dots,10.0$, plotted as functions of the reduced energy $\Delta
E^{(\kappa)}$. Least-sensitive inflection points are marked by dots
and transition energies are indicated by dotted lines.}
\label{fig:micro_largeEk07}
\end{figure}

Consequently, we now perform a systematic microcanonical inflection-point
analysis. The microcanonical entropy and its 
derivatives up to second order in the collapse transition region are
shown in Figure~\ref{fig:micro_largeEk07} as functions of the reduced energy
$\Delta E^{(\kappa)} = E-E_\mathrm{min}^{(\kappa)}$, where 
$E_\mathrm{min}^{(\kappa)}$ is the putative ground-state energy found for a 
polymer with bending stiffness $\kappa$.

The entropy $S$ does not possess any least-sensitive inflection point in this
energy region for $\kappa = 7$. However, we identified a least-sensitive 
inflection
point in the $\beta$ curve at $\Delta E^{(\kappa)} \approx 162$. According to 
our microcanonical inflection-point classification scheme, we consider it  
an
independent second-order transition. It corresponds to a negative maximum in 
the next-higher derivative, 
$\gamma(E)$. Similar to chains with bending stiffness $\kappa \leq 6$, 
only this single
second-order transition occurs in this 
temperature
region.

For slightly increased bending stiffness $\kappa = 7.5$, surprisingly,
least-sensitive inflection points are identified in both $\beta$ and $\gamma$,
in contrast to the expectation of a single transition point from the canonical
analysis. One inflection point in $\beta$ is located at $\Delta E^{(\kappa)}
\approx 158$ and another emerges at the lower energy $\Delta E^{(\kappa)}
\approx 144$ in $\gamma$, suggesting an independent third-order transition
besides the second-order transition.

At the bending stiffness $\kappa = 8.0$, an inflection point at 
$\Delta E^{(\kappa)} \approx 148$ is found in the entropy curve and, thus, 
corresponds to an independent
first-order transition, and associated with it is a 
positive-valued
minimum in $\beta$. This indicates that the extension of the second-order 
collapse transition known from flexible polymers
develops into a first-order transition, whereas the other inflection point in
$\gamma$ at $\Delta E^{(\kappa)} \approx 138$ corresponds to a third-order
transition.

For $\kappa = 8.5$ and $\kappa = 9.0$, least-sensitive inflection
points show up in the entropy curves at 
$\Delta E^{(\kappa)} \approx 142$ and $\Delta E^{(\kappa)} \approx 138$, 
respectively. Therefore,
these transitions are classified as first-order transitions. Interestingly,
identified by the least-sensitive inflection points in $\beta$, the 
transitions
at energies $\Delta E^{(\kappa)} \approx 143$ for $\kappa = 8.5$ and
$\Delta E^{(\kappa)} \approx 127$ for $\kappa = 9.0$, respectively, change to 
second-order
transitions from the third-order signals we found in this transition region 
for $\kappa = 7$ and $\kappa = 7.5$.

Strikingly, for $\kappa = 10.0$, both least-sensitive inflection 
points are found in $S$, which are identified best from the two separate 
positive minima in $\beta$ at $\Delta E^{(\kappa)} \approx 130$ and at 
$\Delta E^{(\kappa)} \approx 110$.
Thus, the new transition branch finally turns into another
first-order transition line.
\begin{figure}
\centering
\includegraphics[width=0.5\textwidth]{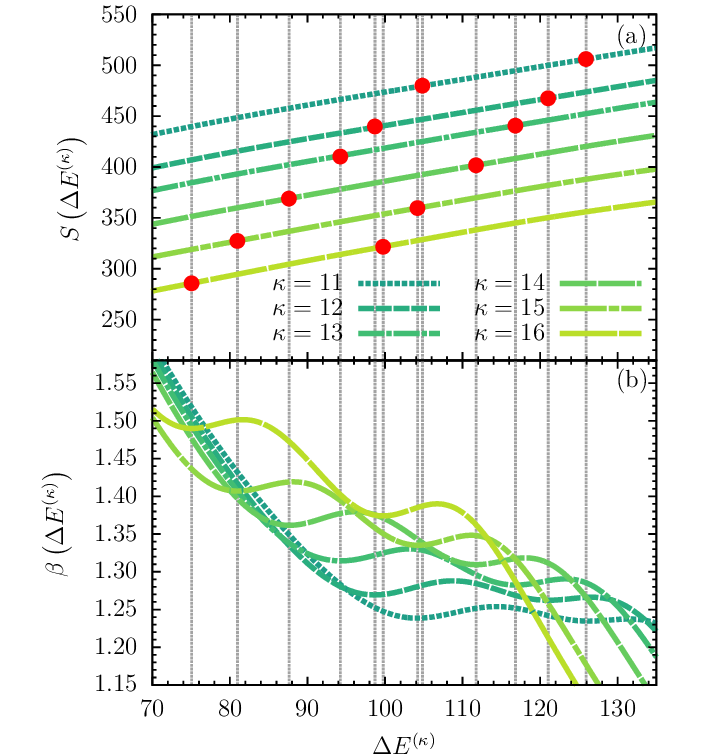}
\caption{(a) Microcanonical entropy $S$ and its derivative (b) $\beta =
dS/dE$ for $\kappa=11,\dots,16$, plotted as
functions of $\Delta E^{(\kappa)}$. Dots show the locations of 
least-sensitive inflection
points; dotted lines were drawn at the transition energies.}
\label{fig:micro_largeEk10}
\end{figure}
\begin{figure*}
\centering
\includegraphics[width=0.80\textwidth]{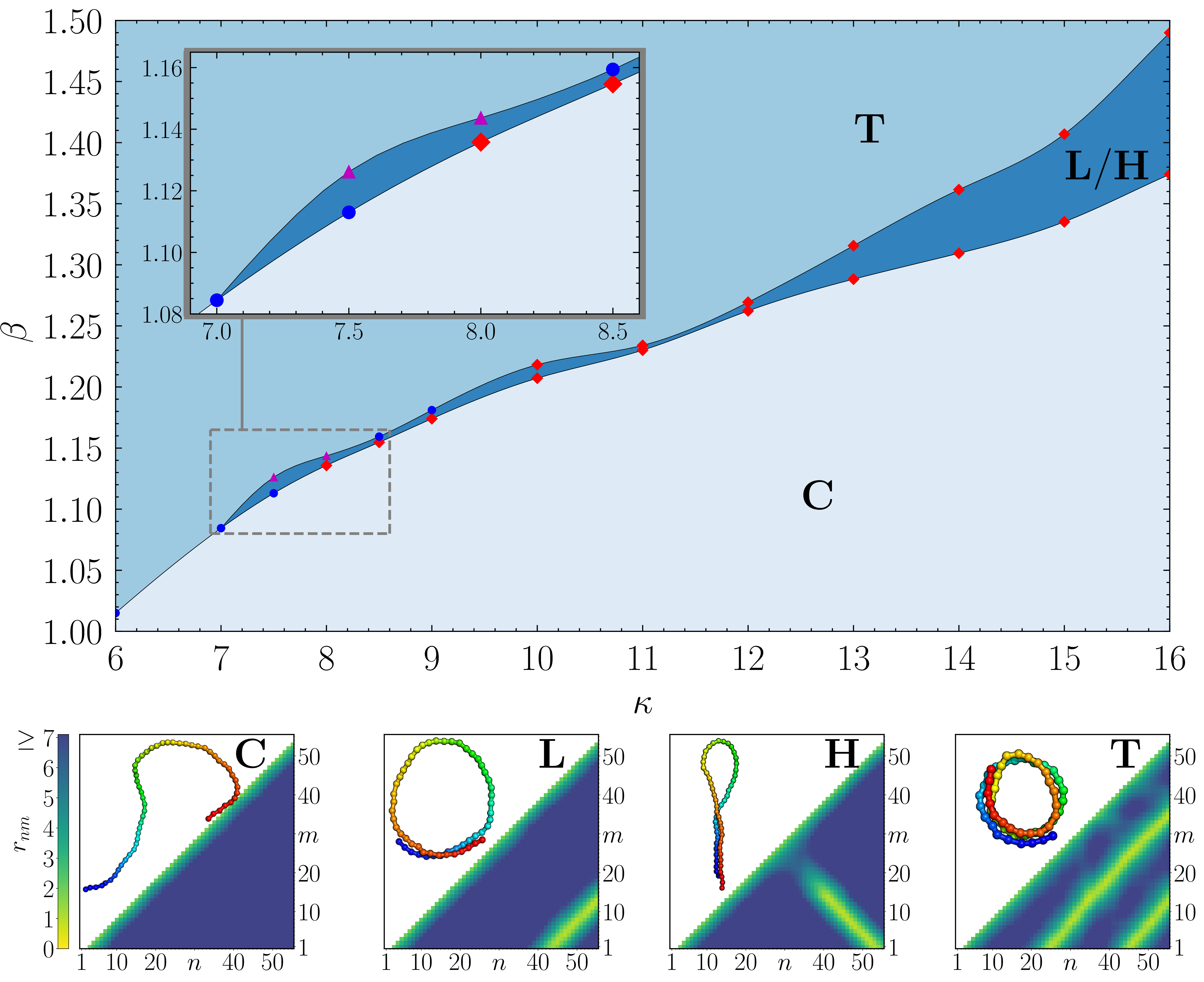}
\caption{Hyperphase diagram for semiflexible polymers with 55 monomers,
parametrized by bending stiffness $\kappa$ and inverse microcanonical
temperature $\beta$. Red diamonds mark first-order, blue dots
second-order, and purple triangles third-order transitions as identified by 
microcanonical inflection-point analysis of results obtained in our 
simulations. Solid
transition lines are guides to the eye. Conformations characteristic for
the respective phases and their distance maps (lower triangles in 
the insets) are shown below the phase diagram. We distinguish 
the following structure types: \textbf{C}:
random coils in the wormlike-chain regime, \textbf{H}: hairpins, \textbf{L}: 
loops, and \textbf{T}: toroids. Monomer labels are ordered from the blue 
(first monomer) to the red end (last monomer).}
\label{fig:micro_phase}
\end{figure*}

Extending the analysis further to bending
stiffness values $\kappa = 11,\dots,16$, we obtain the results shown in
Fig.~\ref{fig:micro_largeEk10}. 
For each $\kappa$ value, only pairs of
least-sensitive inflection points in the entropy curves are found in this 
energy region. Therefore, these are independent first-order transitions that 
are
clearly signaled by positive minima in the $\beta$ curves. As can 
be seen in Fig.~\ref{fig:micro_largeEk10}(a), the transition energy 
difference between
the two first-order transitions increases for stiffer chains. For larger 
bending stiffness values, the
difference of the corresponding microcanonical inverse transition temperatures
is larger as well. The back-bending features in the $\beta$ curves near 
the transitions points are more prominent for
chains with greater bending stiffness. This is an important result as it 
shows that the two transition lines that have formed out of the bifurcation 
point, embrace an entirely new phase and the fact that the first-order 
transition characteristics become more pronounced means that the phase is 
getting more stable as $\kappa$ increases in this region of the phase diagram.

The $\beta-\kappa$ hyperphase diagram constructed from the results we 
obtained by microcanonical inflection-point analysis in the vicinity of the 
bifurcation point is shown in
Fig.~\ref{fig:micro_phase}. 
The extension of the coil-globule transition line remains intact as a
single second-order transition from the flexible case ($\kappa = 0$) up to 
the bifurcation located at about $\kappa = 7$ and $\beta = 1.08$. Note that, 
for bending stiffness values $\kappa>7$, transition types on the upper 
line change from third via second to first order. We have already discussed 
this transition behavior in the context of the microcanonical analysis. 
For finite systems, this is a characteristic feature of transition
lines branching off a main transition line. Transitions of higher-than-second 
order are also
common in finite
systems~\cite{qi2018,qi2019}.
Without their consideration, the phase diagram would contain a gap.

In the higher-temperature regime (low $\beta$), the disordered phase C is
governed by wormlike random-coil structures. In this
regime, entropic effects enable sufficiently large fluctuations that suppress
the formation of stable energetic contacts between monomers. For 
$\kappa \leq 7$, coil structures directly transition
into the toroidal phase T upon lowering the temperature (increasing $\beta$).
However, more interestingly, the formation of a new stable phase between the
random-coil phase C and the toroidal phase T is observed for $\kappa > 7$. 
This
phase is characterized by the coexistence of hairpins (H) and loop (L)
structures. Therefore, we call it a mixed phase. Wormlike chains fold into
hairpins and eventually loops in this transition process.
Further cooling leads to another transition into the toroidal phase
T.

By analyzing the structures in these phases, we found that they possess unique
arrangements of nonbonded monomer-monomer contacts. Therefore, we used 
distance maps of
monomers to characterize these phases. As shown in 
Fig.~\ref{fig:micro_phase}, the
representative conformations as well as their distance maps for each phase are
included. The two nonbonded monomers $n$ and $m$ are considered to be in close
contact if their distance $r_{nm} < 1.2$, but the results are not  
very sensitive
to this choice, as long as counting nonnearest neighbor contacts is 
avoided.
This threshold distance is close to the minimum of the Lennard-Jones 
potential. Nonbonded monomer pairs in contact are
represented by the yellow region in the triangular distance maps shown 
underneath
the conformations. For the extended coil structures, stable contacts  
of monomers
are prevented by thermal fluctuations, and thus they do not exhibit any
particular features. In the intermediate phase, the tails of hairpin  
structures
are in contact, with antiparallel orientation. This results in the contact 
line
that is perpendicular to the diagonal in the contact map, which makes it easy 
to
identify these structures. In loops, on the other hand, the tails align with  
parallel
orientation. Consequently, the short contact line is parallel to the diagonal  
in
the contact map. In contrast to loops, toroids try to reduce system energy by
forming additional contacts. As a result, a second streak parallel to the
diagonal in the contact map accounts for another winding.
\begin{figure}
\centering
\includegraphics[width=0.5\textwidth]{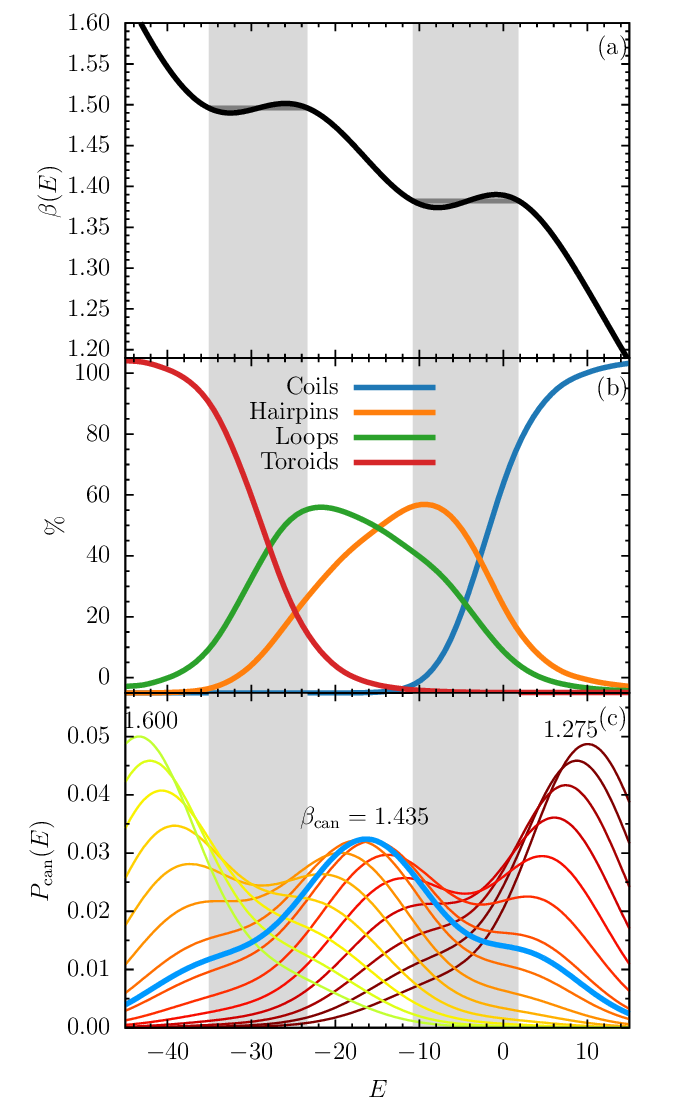}
\caption{(a) Microcanonical temperature and (b) system energy dependence of
frequencies for the different structure types at $\kappa = 16$. (c)
Canonical energy probability distributions $P_{\mathrm{can}}$ at various
inverse thermal energies 
$\beta_{\mathrm{can}} = 1/k_{\mathrm{B}}T_{\mathrm{can}}$.}
\label{fig:mixphase_55}
\end{figure}

In order to quantify the population of different structures in each phase and  
to
gain more insights into the transition behavior in this energy range, we have
also measured the probabilities for each structure type as functions of the 
system energy. Detailed results 
for $\kappa=16$ are presented in Fig.~\ref{fig:mixphase_55}. The $\beta$ 
curve is shown in
Fig.~\ref{fig:mixphase_55}(a) and the frequencies of the
different structure types in Fig.~\ref{fig:mixphase_55}(b). The
microcanonical Maxwell constructions for $\beta$ in the transition regions 
define the
coexistence regions of first-order transitions. These
regions are shaded gray; their widths are a measure for the latent heat. 
There 
is a clearly visible energetic gap between the two transitions (the two 
coexistence regions do not overlap),
confirming that the hairpin-loop crossover is a stable intermediate 
phase. The canonical energy
probability distributions $P_{\mathrm{can}}(E)$ shown in 
Fig.~\ref{fig:mixphase_55}(c) for various canonical
temperatures support this interpretation. There are two noticeable suppressed 
regions in the envelope of these curves, each caused by a
first-order transition. The locations nicely coincide with the 
corresponding first-order
transition regions we identified by microcanonical inflection-point
analysis. Interestingly, the distribution for $\beta_{\mathrm{can}} = 1.435$ 
(blue curve) spans the
entire energy range, and the presence of two transitions is only
reflected
by the two shoulders on either side of the peak. This also helps explain the 
single peak in the specific heat curve for $\kappa=16$ in 
Fig.~\ref{fig:2d-k7-16:can}(c): The fluctuation range covers the entire 
energy region of both transitions. Therefore, the two individual transitions 
cannot be distinguished by the
canonical analysis of response quantities.
\begin{figure}
\centering
\includegraphics[width=1.0\linewidth]{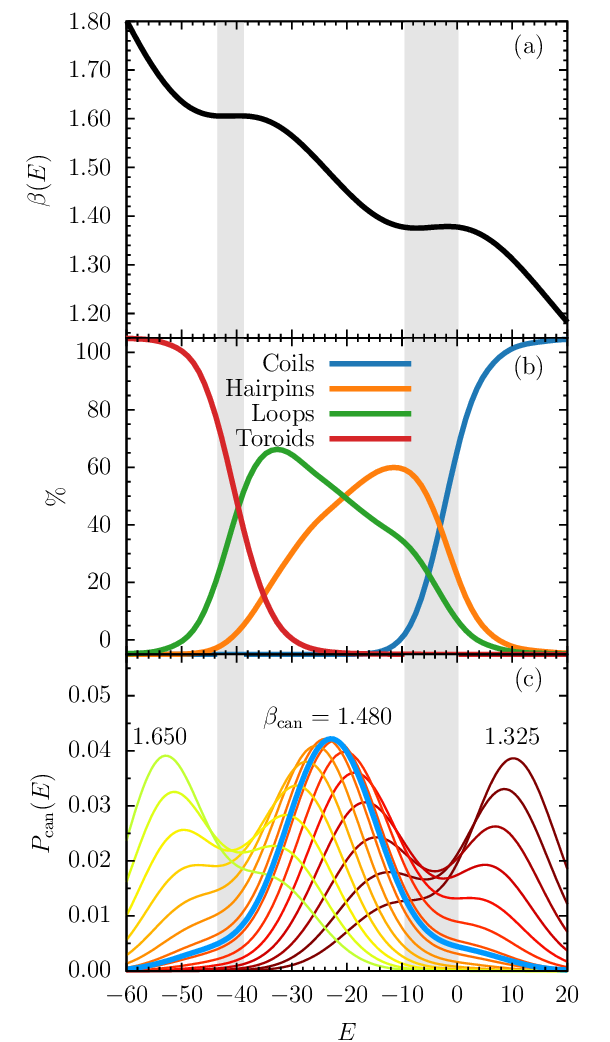}
\caption{Same as Fig.~\ref{fig:mixphase_55}, but for $N=70, \kappa = 30$.}
\label{fig:mixphase_70}
\end{figure}
\begin{figure}
\centering
\includegraphics[width=1.0\linewidth]{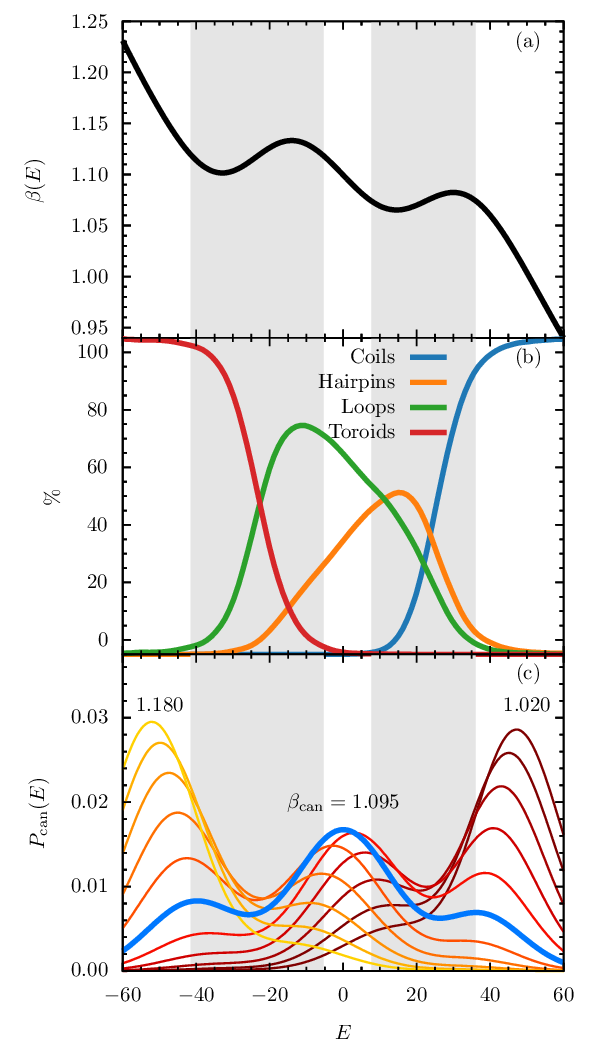}
\caption{Same as Fig.~\ref{fig:mixphase_55}, but for $N=100, \kappa = 45$.}
\label{fig:mixphase_100}
\end{figure}

At high energies, coil structures dominate the phase behavior. Upon lowering 
the energy, there are mainly two ways for the extended
coils to fold and create tail contacts, parallel and antiparallel. During the 
ensuing
transition, the presence of hairpins with antiparallel tail contacts rapidly
increases. These conformations still provide sufficient entropic freedom for 
the
dangling tail, which is already stabilized by van der Waals contacts, however.
The loop part of the hairpin helps reduce the stiffness restraint. It is
noteworthy that the pure loop structures with parallel tails also 
significantly
contribute to the population, although at a lesser scale in this region. The
actual crossover from hairpins to loops happens within the phase. For
lower energies in this mixed phase, the population of hairpins decreases,
whereas loops take over dominance. Moreover, the energy difference between 
these
two types of structures is sufficiently small for thermal fluctuations
to easily convert one structure type to the other by folding the tails back 
to or
away from the loop part. This also explains why there is no
phase transition between them. Importantly, even though hairpin and loop
structures are irrelevant at very low temperatures, they are
biologically significant secondary structure types at finite temperatures. The
tail can be easily spliced, contact pair by contact pair, with little 
energetic
effort, which supports essential micromolecular processes on the DNA and RNA
level such as transcription and translation. Therefore, it is important to
discern the phase dominated by these structures.

Upon further reducing the energy (and therefore also entropy), forming
energetically favorable van der Waals contacts becomes the dominant structure
formation strategy and loops coil in to eventually form toroids. In contrast 
to
loops, toroidal structures are more ordered and stabilized further by
additional energetically favorable attractions between monomers.

To test the robustness of the obtained results, we have also performed  
selected
simulations of semiflexible chains of this generic model with 70 and 100
monomers, essentially yielding the same qualitative results. 
Most importantly, the
bifurcation of the collapse transition line is also observed for these chain 
lengths.
Quantitatively, the bifurcation point is shifted toward higher 
bending
stiffness. For $N=70$, it is located at about $\kappa = 20$, 
whereas it is near $\kappa = 40$ for $N=100$. This is expected, of course,
as the number of possible energetic contacts scales with the number 
of monomers,
which requires a larger energy penalty to break these contacts. It also helps
understand why microbiological structures are not only finite but exist on a
comparatively small, mesoscopic length scale. At the physiological scale,
structure formation processes of large systems would be much more difficult to
control and stabilize. This also means that studying such systems in the
thermodynamic limit may not actually aid in understanding physics at 
mesoscopic scales.

We also analyzed the 
structure population for the two first-order transitions at these 
chain lengths. The results for ($N=70, \kappa = 30$) and ($N=100, \kappa = 
45$) are
shown in Fig.~\ref{fig:mixphase_70} and Fig.~\ref{fig:mixphase_100}, 
respectively. In both cases the same crossover
behavior of hairpins and loops in the intermediate phases as for 
$N=55$ is observed. The two
suppressed regions of canonical energy distribution probabilities are clearly
visible as well. These results support the generality of our conclusions for 
semiflexible polymers on mesoscopic scales.
\section{Summary}
\label{sec:sum}
We employed the microcanonical inflection-point analysis 
method
to study the transition behavior of a generic semiflexible polymer model with 
self-interactions. 
The replica-exchange Monte Carlo simulation method, extended to cover both 
temperature 
and bending stiffness parameter spaces, was used to 
obtain highly accurate estimates of the density of states needed for the 
microcanonical
statistical analysis. Advanced structural Monte Carlo updates helped improve 
the 
efficiency of our simulations. Least-sensitive inflection points in the
microcanonical entropy and its derivatives were used as indicators for 
the systematic identification and classification of phase
transitions. 

The coarse-grained semiflexible polymer we mainly studied consists of 55 
repetitive units 
(monomers). This chain has been extensively studied in the flexible limit. 
Therefore, in this study, we focused on the transition line that extends from 
the collapse transition point for flexible polymers upon increasing the value 
of the bending stiffness parameter of the chain. Below a certain threshold, 
for nonzero bending stiffness, the line separates wormlike-chain coil 
structures from 
toroidal conformations. However, remarkably, we find that  
the line bifurcates at a certain bending strength, which leads to the  
formation of an 
intermediate, mixed phase of stable secondary structures such as hairpins 
and loops. Whereas the extended collapse transition line below the 
bifurcation point is of second order, the two transitions embracing the novel 
intermediate phase eventually turn into comparatively strong first-order 
transitions. In the bifurcation region, the new transition branch starts off 
with third-order, then second-order, and finally first-order transition 
behavior for sufficiently large values of the bending stiffness.

It is worth emphasizing that these transition 
signals are indistinguishable in canonical statistical analysis, where the 
energetic
fluctuations are large enough to envelop both transitions. 
Therefore, conventional 
canonical statistical analysis of 
fluctuations of response quantities was found to be too insensitive to reveal 
the separate phase of microbiologically important secondary structures.

As detailed structural analysis showed, the mixed intermediate phase is 
populated by both loop
and hairpin structures, which are biologically relevant secondary structures 
present in 
DNA and RNA under physiological conditions. These structures are 
inherently finite in size. Yet, studies of chains with different lengths also 
confirmed that our results are robust and this intermediate phase is stable. 

Our results support the conclusion that biomolecular function is inevitably 
connected to structural features of segments of macromolecules on 
mesoscopic scales. For semiflexible polymers, hairpins and loops are the 
optimal secondary structure types that represent the best compromise of 
structural stability and variability under thermal conditions. Stability is 
achieved by reducing energy and variability by increasing entropy. Therefore, 
biologically relevant functional polymers must be able to adapt to 
environmental conditions by resisting random energetic 
fluctuations, but also allowing for structural changes on small energetic
scales to ensure predictable functional behavior. Therefore, as our study 
showed, only semiflexible polymers with sufficient bending stiffness 
exceeding a certain threshold can form a separate intermediate phase 
of secondary structures, making them excellent candidates for functional 
biomolecules.  
\section{Conflicts of Interest}
There are no conflicts of interest to declare.
\section{Acknowledgment}
This study was supported in part by resources and technical expertise from the
Georgia Advanced Computing Resource Center (GACRC).

\begin{thebibliography}{99}
%
\bibitem{ferreira2015}
Ferreira, L. G.; Dos Santos, R. N.; Oliva, G.; Andricopulo, A. D. Molecular 
Docking and Structure-Based Drug Design Strategies. \textit{Molecules} 
\textbf{2015}, 20 (7), 13384-13421.
%
\bibitem{gao2021}
Gao, P.; Nicolas, J.; Ha-Duong, T. Supramolecular Organization of Polymer 
Prodrug Nanoparticles Revealed by Coarse-Grained Simulations. \textit{J. Am. 
Chem. Soc.} \textbf{2021}, 143 (42), 17412-17423.
%
\bibitem{schmidt2014}
Schmidt, T.; Bergner, A.; Schwede, T. Modelling Three-Dimensional Protein 
Structures for Applications in Drug Design. \textit{Drug Discov. Today} 
\textbf{2014}, 19 (7), 890-897.
%
\bibitem{smiatek2020}
Smiatek, J.; Jung, A.; Bluhmki, E. Towards a Digital Bioprocess Replica: 
Computational Approaches in Biopharmaceutical Development and Manufacturing. 
\textit{Trends Biotechnol.} \textbf{2020}, 38 (10), 1141-1153. 
%
\bibitem{kmiecik2016}
Kmiecik, S.; Gront, D.; Kolinski, M.; Wieteska, L.; Dawid, A. E.; Kolinski, A.
Coarse-Grained Protein Models and Their Applications. \textit{Chem. Rev.} 
\textbf{2016}, 116 (14),
7898-7936. 
%
\bibitem{singh2019}
Singh, N.; Li, W. Recent Advances in Coarse-Grained Models for Biomolecules 
and Their Applications.\textit{ Int. J. Mol. Sci.} \textbf{2019}, 20 (15), 
3774. 
%
\bibitem{bachmann2014}
Bachmann, M. \textit{Thermodynamics and Statistical Mechanics of 
Macromolecular Systems}; Cambridge University Press, 2014.
%
\bibitem{yuan2008}
Yuan, C.; Chen, H.; Lou, X. W.; Archer, L. A. DNA Bending Stiffness on Small 
Length Scales. \textit{Phys. Rev. Lett.} \textbf{2008}, 100 (1), 018102.
%
\bibitem{li2018}
Li, S.; Erdemci-Tandogan, G.; van der Schoot, P.; Zandi, R. The Effect of RNA
Stiffness on the Self-Assembly of Virus Particles. J. Phys.: \textit{Condens.
Matter} \textbf{2018}, 30 (4), 044002.
%
\bibitem{kratky1949}
Kratky, O.; Porod, G. R\"ontgenuntersuchung Gel\"oster Fadenmolek\"ule. 
\textit{Recl. Trav. Chim. Pays-Bas} \textbf{1949}, 68 (12), 1106-1122.
%
\bibitem{seaton2013}
Seaton, D. T.; Schnabel, S.; Landau, D. P.; Bachmann, M. From Flexible to 
Stiff: Systematic Analysis of Structural Phases for Single Semiflexible 
Polymers. \textit{Phys. Rev. Lett.} \textbf{2013}, 110 (2), 028103.
%
\bibitem{marenz2016}
Marenz, M.; Janke, W. Knots as a Topological Order Parameter for Semiflexible 
Polymers. \textit{Phys. Rev. Lett.} \textbf{2016}, 116 (12), 128301. 
%
\bibitem{skrbic2016}
Skrbic, T.; Hoang, T. X.; Giacometti, A. Effective Stiffness and Formation of 
Secondary Structures in a Protein-like Model. \textit{J. Chem. Phys.} 
\textbf{2016}, 145 (8), 084904.
%
\bibitem{wu2018}
Wu, J.; Cheng, C.; Liu, G.; Zhang, P.; Chen, T. The Folding Pathways and 
Thermodynamics of Semiflexible Polymers. \textit{J. Chem. Phys.} 
\textbf{2018}, 148 (18), 184901.
%
\bibitem{wu2018b}
Wu, J.; Huang, Y.; Yin, H.; Chen, T. The Role of Solvent Quality and Chain 
Stiffness on the End-to-End Contact Kinetics of Semiflexible Polymers. 
\textit{J. Chem. Phys.} \textbf{2018}, 149 (23), 234903.
%
\bibitem{ab1}
Aierken, D.; Bachmann, M. Comparison of Conformational Phase Behavior for
Flexible and Semiflexible Polymers. \textit{Polymers} \textbf{2020}, 12, 3013.
%
\bibitem{majumder2021}
Majumder, S.; Marenz, M.; Paul, S.; Janke, W. Knots Are Generic Stable Phases 
in Semiflexible Polymers. \textit{Macromolecules} \textbf{2021}, 54(12), 
5321-5334.
%
\bibitem{walker2022}
Walker, C. C.; Fobe, T. L.; Shirts, M. R. How Cooperatively Folding Are 
Homopolymer Molecular Knots? \textit{Macromolecules} \textbf{2022}, 55 (19), 
8419-8437. 
%
\bibitem{shakirov2023}
Shakirov, T.; Paul, W. Aggregation and Crystallization of Small Alkanes. 
\textit{J. Chem. Phys.} \textbf{2023}, 158 (9), 094905.
%
\bibitem{aierken2023}
Aierken, D.; Bachmann, M. Stable Intermediate Phase of Secondary Structures 
for Semiflexible Polymers. \textit{Phys. Rev. E} \textbf{2023}, 107 (3), 
L032501.
%
\bibitem{ab3}
Aierken, D.; Bachmann, M. Impact of Bending Stiffness on Ground State 
Conformations for Semiflexible Polymers. \textit{J. Chem. Phys.} 
\textbf{2023}, 158 (21), 214905.
%
\bibitem{qi2018}
Qi, K.; Bachmann, M. Classification of Phase Transitions by Microcanonical
Inflection-Point Analysis. \textit{Phys. Rev. Lett.} \textbf{2018}, 120 (18),
180601.
%
\bibitem{bird1987}
Bird, R. B.; Armstrong, R. C.; Hassager, O. Dynamics of Polymeric Liquids. 
\textit{Vol. 1: Fluid Mechanics.} \textbf{1987}.
%
\bibitem{kremer1990}
Kremer, K.; Grest, G. S. Dynamics of Entangled Linear Polymer Melts: A 
Molecular-dynamics Simulation. \textit{J. Chem. Phys.} \textbf{1990}, 92 (8), 
5057-5086.
%
\bibitem{milchev2001}
Milchev, A.; Bhattacharya, A.; Binder, K. Formation of Block Copolymer 
Micelles in Solution: A Monte Carlo Study of Chain Length Dependence.
\textit{Macromolecules} \textbf{2001}, 34 (6), 1881-1893.
%
\bibitem{qi2014}
Qi, K.; Bachmann, M. Autocorrelation Study of the $\Theta$ Transition for a
Coarse-Grained Polymer Model. \textit{J. Chem. Phys.} \textbf{2014}, 141 (7),
074101. 
%
\bibitem{qi2019}
Qi, K.; Liewehr, B.; Koci, T.; Pattanasiri, B.; Williams, M. J.; Bachmann, M. 
Influence of Bonded Interactions on Structural Phases of Flexible Polymers. 
\textit{J. Chem. Phys.} \textbf{2019}, 150 (5), 054904.
%
\bibitem{swendsen1986}
Swendsen, R. H.; Wang, J.-S. Replica Monte Carlo Simulation of Spin-Glasses. 
\textit{Phys. Rev. Lett.} \textbf{1986}, 57 (21), 2607.
%
\bibitem{ferrenberg1988}
Ferrenberg, A. M.; Swendsen, R. H. New Monte Carlo Technique for Studying 
Phase Transitions. \textit{Phys. Rev. Lett.} \textbf{1988}, 61 (23), 2635.
%
\bibitem{hukushima1996}
Hukushima, K.; Nemoto, K. Exchange Monte Carlo Method and Application to Spin 
Glass Simulations. \textit{J. Phys. Soc. Jpn.} \textbf{1996}, 65 (6), 
1604-1608.
%
\bibitem{hukushima1996b}
Hukushima, K.; Takayama, H.; Nemoto, K. Application of an Extended Ensemble 
Method to Spin Glasses. \textit{Int. J. Mod. Phys. C} \textbf{1996}, 07, 07, 337-344
%
\bibitem{earl2005}
Earl, D. J.; Deem, M. W. Parallel Tempering: Theory, Applications, and New 
Perspectives. \textit{Phys. Chem. Chem. Phys.} \textbf{2005}, 7 (23), 3910.
%
\bibitem{geyer1991}
Geyer, C. J. Computing Science and Statistics: Proceedings of the 23rd 
Symposium on the Interface. \textit{J. Am. Stat. Assoc.} \textbf{1991}, 156.
%
\bibitem{fiore2008}
Fiore, C. E. First-Order Phase Transitions: A Study through the Parallel 
Tempering Method. \textit{Phys. Rev. E} \textbf{2008}, 78 (4), 041109.
%
\bibitem{kofke2004}
Kofke, D. A. Comment on ``The Incomplete Beta Function Law for Parallel 
Tempering Sampling of Classical Canonical System'' [J. Chem. Phys. 120, 4119 
(2004)]. \textit{J. Chem. Phys.} \textbf{2004}, 121 (2), 1167-1167.
%
\bibitem{machta2009}
Machta, J. Strengths and Weaknesses of Parallel Tempering. \textit{Phys. Rev. 
E} \textbf{2009}, 80 (5), 056706.
%
\bibitem{machta2011}
Machta, J.; Ellis, R. S. Monte Carlo Methods for Rough Free Energy Landscapes: 
Population Annealing and Parallel Tempering. \textit{J. Stat. Phys.} 
\textbf{2011}, 144 (3), 541-553.
%
\bibitem{predescu2005}
Predescu, C.; Predescu, M.; Ciobanu, C. V. On the Efficiency of Exchange in 
Parallel Tempering Monte Carlo Simulations. \textit{J. Phys. Chem. B} 
\textbf{2005}, 109 (9), 4189-4196.
%
\bibitem{schnabel2011b}
Schnabel, S.; Janke, W.; Bachmann, M. Advanced Multicanonical Monte Carlo
Methods for Efficient Simulations of Nucleation Processes of Polymers.
\textit{J. Comput. Phys.} \textbf{2011}, 230 (12), 4454-4465.
%
\bibitem{williams2016}
Williams, M. J.; Bachmann, M. Significance of Bending Restraints for the 
Stability of Helical Polymer Conformations. \textit{Phys. Rev. E} 
\textbf{2016}, 93 (6), 062501. 
%
\bibitem{austin2018a}
Austin, K. S.; Marenz, M.; Janke, W. Efficiencies of Joint Non-Local Update 
Moves in Monte Carlo Simulations of Coarse-Grained Polymers. \textit{Comput. 
Phys. Commun.} \textbf{2018}, 224, 222-229.
%
\bibitem{hukushima1999}
Hukushima, K. Domain-Wall Free Energy of Spin-Glass Models: Numerical Method 
and Boundary Conditions. \textit{Phys. Rev. E} \textbf{1999}, 60 (4), 
3606-3613.
%
\bibitem{rozada2019}
Rozada, I.; Aramon, M.; Machta, J.; Katzgraber, H. G. Effects of Setting 
Temperatures in the Parallel Tempering Monte Carlo Algorithm. \textit{Phys. 
Rev. E} \textbf{2019}, 100 (4), 043311.
%
\bibitem{kumar1992a}
Kumar, S.; Rosenberg, J. M.; Bouzida, D.; Swendsen, R. H.; Kollman, P. A. The 
Weighted Histogram Analysis Method for Free-energy Calculations on 
Biomolecules. I. The Method. \textit{J. Comput. Chem.} \textbf{1992}, 13 (8), 
1011-1021.
%
\bibitem{bezier1968}
B\'ezier, P. Proc\'ed\'e de D\'efinition Num\'erique Des Courbes et Surfaces 
Non Math\'ematiques. \textit{Automatisme} \textbf{1968}, 13 (5), 189-196.
%
\bibitem{gordon1974}
Gordon, W. J.; Riesenfeld, R. F. Bernstein-B\'ezier Methods for the 
Computer-Aided Design of Free-Form Curves and Surfaces. \textit{J. ACM} 
\textbf{1974}, 21 (2), 293-310.
%
\bibitem{gross2001}
Gross, D. H. E. Microcanonical Thermodynamics: \textit{Phase Transitions in 
''Small'' Systems}; World Scientific, 2001.
%
\bibitem{stevenson1981}
Stevenson, P. M. Resolution of the Renormalisation-Scheme Ambiguity in 
Perturbative QCD. \textit{Phys. Lett. B} \textbf{1981}, 100 (1), 61-64.
%
\bibitem{stevenson1981b}
Stevenson, P. M. Optimized Perturbation Theory. \textit{Phys. Rev. D} 
\textbf{1981}, 23 (12), 2916-2944.
%
\bibitem{sitarachu2020}
Sitarachu, K.; Bachmann, M. Phase Transitions in the Two-Dimensional Ising
Model from the Microcanonical Perspective. \textit{J. Phys.: Conf. Ser.}
\textbf{2020}, 1483, 012009.
%
\bibitem{sitarachu2020b}
Sitarachu, K.; Zia, R. K. P.; Bachmann, M. Exact Microcanonical Statistical 
Analysis of Transition Behavior in Ising Chains and Strips. \textit{J. Stat. 
Mech.} \textbf{2020}, 2020 (7), 073204. 
%
\bibitem{sitarachu2022}
Sitarachu, K.; Bachmann, M. Evidence for Additional Third-Order Transitions in 
the Two-Dimensional Ising Model. \textit{Phys. Rev. E} \textbf{2022}, 106 (1), 
014134.
%
\bibitem{trugilho2022}
Trugilho, L. F.; Rizzi, L. G. Microcanonical Characterization of First-Order 
Phase Transitions in a Generalized Model for Aggregation. \textit{J. Stat. 
Phys.} \textbf{2022}, 186 (3), 40.
%
\bibitem{trugilho2022b}
Trugilho, L. F.; Rizzi, L. G. Shape-Free Theory for the Self-Assembly Kinetics
in Macromolecular Systems. \textit{EPL} \textbf{2022}, 137 (5), 57001.
%
\bibitem{bel-hadj-aissa2020}
Bel-Hadj-Aissa, G.; Gori, M.; Penna, V.; Pettini, G.; Franzosi, R. Geometrical 
Aspects in the Analysis of Microcanonical Phase-Transitions. \textit{Entropy} 
\textbf{2020}, 22 (4), 380.
%
\bibitem{gori2022}
Gori, M.; Franzosi, R.; Pettini, G.; Pettini, M. Topological Theory of Phase 
Transitions. \textit{J. Phys. A: Math. Theor.} \textbf{2022}, 55 (37), 375002.
%
\bibitem{pettini2019}
Pettini, G.; Gori, M.; Franzosi, R.; Clementi, C.; Pettini, M. On the Origin 
of Phase Transitions in the Absence of Symmetry-Breaking. \textit{Phys. A: 
Stat. Mech.} \textbf{2019}, 516, 376-392.
%
\bibitem{dicairano2022}
Di Cairano, L.; Capelli, R.; Bel-Hadj-Aissa, G.; Pettini, M. Topological 
Origin of the Protein Folding Transition. \textit{Phys. Rev. E} \textbf{2022}, 
106 (5), 054134.
%
\bibitem{bel-hadj-aissa2020b}
Bel-Hadj-Aissa, G. High Order Derivatives of Boltzmann Microcanonical Entropy 
with an Additional Conserved Quantity. \textit{Phys. Lett. A} \textbf{2020}, 
384 (24), 126449. 
%
\bibitem{chaudhuri2021}
Chaudhuri, A.; Sadek, C.; Kakde, D.; Wang, H.; Hu, W.; Jiang, H.; Kong, S.; 
Liao, Y.; Peredriy, S. The Trace Kernel Bandwidth Criterion for Support Vector 
Data Description. \textit{Pattern Recognit.} \textbf{2021}, 111, 107662.
%
\end{thebibliography}
\end{document}